\pdfoutput=1

\documentclass[11pt,twoside,a4paper,cmspaper,final,collab]{cms-tdr}
\def\svnVersion{315673}\def\cmsCernNo{2013-037}\def\cmsCernDate{\today}\def\cmsMessage{Published in Physical Review D as \href{http://dx.doi.org/10.1103/PhysRevD.92.112001}{\doi{10.1103/PhysRevD.92.112001.}}}

\begin{document}\cmsNoteHeader{FSQ-12-032}

\hyphenation{had-ron-i-za-tion}
\hyphenation{cal-or-i-me-ter}
\hyphenation{de-vices}
\RCS$Revision: 309372 $
\RCS$HeadURL: svn+ssh://svn.cern.ch/reps/tdr2/papers/FSQ-12-032/trunk/FSQ-12-032.tex $
\RCS$Id: FSQ-12-032.tex 309372 2015-11-07 08:55:40Z jung $

\newcommand{\ptmin}{\ensuremath{p^{\mathrm{min}}_{\mathrm{T}}}\xspace}
\newcommand{\ptlead}{\ensuremath{p^{\mathrm{lead}}_{\mathrm{T}}}\xspace}
\newcommand{\ptz}{\ensuremath{p_{\mathrm{T,0}}}\xspace}
\providecommand{\EPOS}{\textsc{epos}\xspace}
\newlength\cmsFigWidth
\ifthenelse{\boolean{cms@external}}{\setlength\cmsFigWidth{0.85\columnwidth}}{\setlength\cmsFigWidth{0.4\textwidth}}
\ifthenelse{\boolean{cms@external}}{\providecommand{\cmsLeft}{top}}{\providecommand{\cmsLeft}{left}}
\ifthenelse{\boolean{cms@external}}{\providecommand{\cmsRight}{bottom}}{\providecommand{\cmsRight}{right}}
\cmsNoteHeader{AN-13-103}
\title{Production of leading charged particles and leading charged-particle jets at small transverse momenta in pp collisions at \texorpdfstring{$\sqrt{s}=8$\TeV}{sqrt(s)=8 TeV}}

\date{\today}

\abstract{
The per-event yield of the highest transverse momentum charged particle and charged-particle jet, integrated
 above a given \ptmin  threshold starting at $\ptmin =0.8$ and 1\GeV, respectively, is studied in $\Pp\Pp$ collisions at $\sqrt{s} = 8$\TeV.
The particles and the jets are measured in the pseudorapidity ranges
$\abs{\eta}< 2.4$ and 1.9, respectively. The data are sensitive to the momentum scale at which
parton densities saturate in the proton, to multiple partonic interactions, and other key
aspects of the transition between the soft and hard QCD regimes in hadronic collisions.
}

\hypersetup{%
pdfauthor={CMS Collaboration},%
pdftitle={Production of leading charged particles and leading charged-particle jets at small transverse momenta in pp collisions at sqrt(s)=8 TeV},%
pdfsubject={CMS},%
pdfkeywords={CMS, inclusive cross section}}

\maketitle

\section{Introduction}
\label{sec:intro}

The production of jets with large transverse momenta $\pt \gg \Lambda_{\mathrm{QCD}} \approx 0.2\GeV$ in high-energy proton-proton (pp) collisions originates from the scattering of partons, a process described by perturbative quantum chromodynamics (pQCD), through the convolution of the parton-parton cross section with the density of partons inside the protons.
Jet production in $\Pp\Pp$ collisions at the LHC, at transverse momenta $\pt > 20\GeV$ and in the pseudorapidity range $\abs{\eta} < 3$, is well described by next-to-leading-order pQCD calculations~\cite{Aad:2011fc, Chatrchyan:2012bja, CMS:2011ab}. However,
most of the final-state hadrons produced in $\Pp\Pp$ collisions arise from the hadronisation
of quarks and gluons scattered through ``semi-hard" interactions with exchanged momenta of
$\mathcal{O}$(1--3\GeV). At such low values of \pt, the theoretical partonic cross section, $\rd\sigma/\rd\pt^2 \propto \alpha_S^2 (\pt)/\pt^4$, where $\alpha_S$ is the strong coupling,
becomes very large, and the integrated cross section
$\sigma(\ptmin) = \int_{\ptmin}  \rd\pt^2 d\sigma/\rd\pt^2 $ exceeds the total inelastic $\Pp\Pp$ cross section, $\sigma_\text{inel}$.
At $\sqrt{s} = 8$\TeV, where $\sigma_\text{inel} \approx 70$\unit{mb}~\cite{Antchev:2013paa}, this occurs
at \ptmin values of $\mathcal{O}(3\GeV)$, much larger than the QCD scale, $\Lambda_{\mathrm{QCD}}$, at which the strong coupling diverges~\cite{Sjostrand:1987su,Sjostrand:2004pf}.

Model calculations of hadronic collisions often regulate such an infrared divergence through an effective parameter connected to the confinement scale of hadrons~\cite{Dischler:2000pk}, such that the leading particle or leading jet production cross sections do not exceed the value of $\sigma_\text{inel}$. Contrary to the inclusive particle or jet production cross sections, the \textit{leading} particle or \textit{leading} jet production cross sections must indeed approach the total inelastic cross section because only one particle or one jet, the one with highest \pt in this case, is considered per event. In addition, at small \pt, the parton densities are probed in a region where parton recombination, \ie saturation (see \eg Ref.~\cite{Mueller:1989st}), may occur.

Reference~\cite{Grebenyuk:2012qp} proposes that the jet cross section integrated over $\pt > \ptmin$ can be used as a probe of the transition from the perturbative ($\ptmin \gg \Lambda_{\mathrm{QCD}}$) to the non-perturbative region ($\ptmin \to \Lambda_{\mathrm{QCD}}$). According to Ref.~\cite{Grebenyuk:2012qp}, this
transition should also be visible for cross sections defined in restricted ranges of pseudorapidity.

The results presented in this paper are based on measurements of single charged particles and jets reconstructed from charged particles alone.
The advantage of jets is that they include more particles originating from the outgoing partons, while single charged hadrons carry only a fraction of the parent parton momentum. On the other hand, jets are sensitive to the underlying event (UE) activity,
consisting of particles originating from multiple partonic interactions (MPI) and initial and final state radiation, while single leading tracks are not. The measurements based on leading particles and leading jets are therefore complementary.
Throughout the text, the term ``track-jets" refers to detector-level jets, reconstructed from charged-particle tracks observed in the detector, while ``charged-particle jets" or just ``jets" denote corrected, stable-particle level jets, consisting of stable charged particles from the final state.

In this paper, the yields, $r(\ptmin)$, for $\Pp\Pp$ collisions with a leading charged particle or a leading jet are measured as a function of a minimum transverse momentum, \ptmin:
\begin{equation}
r(\ptmin)
= \frac{1}{N_{\text{evt}}}  \int_{\ptmin} \rd\ptlead \left( \frac{\rd{}N}{\rd\ptlead}\right),
\label{int}
\end{equation}
where $N_{\text{evt}}$ is the number of selected events with a leading charged particle with $\pt > 0.4\GeV$ and $\abs{\eta} < 2.4$, and $N$ is the number of events with a leading charged particle or a leading jet with transverse momentum \ptlead within  $\abs{\eta} < 2.4$ or 1.9, respectively.

\section{Phenomenological models}
The measured distributions are compared to the predictions of different hadronic interaction
models whose tunable parameters (mostly connected to non-perturbative and semi-hard QCD phenomena)
are obtained from comparisons to LHC data such as those on UE activity, inclusive
multiparticle production and diffraction.

The \PYTHIA6~\cite{Sjostrand:2006za} and 8~\cite{Sjostrand:2007gs} event generators tame the low-\pt behaviour of the leading-order
pQCD $2{\to}2$ cross sections with a phenomenological factor \cite{Sjostrand:1987su,Sjostrand:2004pf} $[\alpha_s^2(\ptz^2 + \pt^2)/\alpha_s^2(\pt^2)] \linebreak[2][{\pt^4}/{(\ptz^2 + \pt^2)^2}]$, where $\ptz$ is a (tunable) infrared regulator that runs with centre-of-mass energy.
The tunes 4C~\cite{Corke:2010yf}, CUET~\cite{Gunnellini:2014zba}, and MONASH~\cite{Skands:2014pea} are used, featuring different choices of the $\ptz$ cutoff, proton transverse profile, and/or parton distribution functions.

The {\HERWIGpp} \cite{Bahr:2008pv} Monte Carlo (MC) includes a hard (pQCD $2{\to}2$ interactions)~\cite{Bahr:2008dy} and a soft (non-perturbative) component for multiple interactions~\cite{Herwig:2009soft}. The soft part is parametrised phenomenologically as $\rd\sigma/\rd{}\pt^2=A{\re}^{-\beta \pt^2}$. The transition scale between the hard and the soft regions is set by the parameter $\ptz$, obtained from fits to MPI and UE data, as well as to the effective cross section for double-parton scatterings. The parameters $A$ and $\beta$ are fixed by the requirements that the transverse momentum distribution be continuous at the matching scale $\ptz$, and that the model reproduces the measured total cross section. {\HERWIGpp} with tune  CUETHS1 \cite{GEN-14-001}  is used to compared the measurement at reconstruction level. {\HERWIGpp} is not used in the comparison of the final results, as it does not contain any diffractive component.

The other two models, \textsc{qgsjet}-II~\cite{Ostapchenko:2010vb} and \EPOS~\cite{Werner:2005jf,Pierog:2013ria}, are based on
Regge-Gribov effective field
theory~\cite{Gribov:1968fc}, which allows for a consistent treatment of soft and
hard scattering processes in terms of the same degrees of freedom (reggeons and pomerons), based on
unitarity cuts of the corresponding elastic scattering diagrams.
Perturbative parton-parton
processes are obtained via ``cut (hard) pomeron'' diagrams, and multi-scattering phenomena
(saturation, MPI) are implemented through various procedures~\cite{dEnterria:2011kw}. The two models differ in
their approximations for the collision configurations (with exact energy sharing imposed in the case of
\EPOS), and the treatment of diffractive and perturbative contributions (the effective soft-hard
transition occurs at $\ptz\sim1.6$\GeV for \textsc{qgsjet}-II and at $\ptz\sim2$\GeV for
\EPOS). Finally, in
contrast to other MCs,
\textsc{epos} includes also collective expansion effects in the final state that boost
the final \pt distribution of the produced hadrons. It is worth to highlight that, for all MC models,
the (centre-of-mass energy dependent) $\ptz$ cutoff plays a very similar role to the ``saturation scale''
($Q_\text{sat}$), which controls the onset of gluon fusion effects in the parton densities~\cite{Gribov:1984tu}.

\section{Experimental analysis}
\label{sec:detector}

The central feature of the CMS apparatus is a superconducting solenoid of 6\unit{m} internal diameter, providing a magnetic field of 3.8\unit{T}. A silicon pixel and strip tracker, a lead tungstate crystal electromagnetic calorimeter, and a brass and scintillator sampling hadron calorimeter are located within the volume of the solenoid.

The inner silicon tracker measures charged-particle trajectories (``tracks" in the following) within the pseudorapidity range $\abs{\eta} < 2.5$. It provides an impact parameter resolution of about 100$\mum$ and a \pt resolution of about 0.7\% for 1\GeV tracks at $\eta = 0$~\cite{Chatrchyan:2014fea}. A more detailed description of the CMS detector, together with definitions of the coordinate system and kinematic variables can be found in Ref.~\cite{Chatrchyan:2008zzk}.

The data analysed in this study were collected during a dedicated proton-proton run with an integrated luminosity of 45\mubinv at a centre-of-mass energy of $\sqrt{s}=8$\TeV. This run has a low instantaneous luminosity and a low probability ($\sim$2\%) of multiple $\Pp\Pp$ interactions occurring in the same bunch crossing (pileup). Pileup events are rejected by requiring exactly one vertex, following the method described in Ref.~\cite{Chatrchyan:2014qka}.

Minimum bias events were selected online with the TOTEM T2 telescopes~\cite{totem} that are placed symmetrically at about 14\unit{m} on both sides from the interaction point (IP). 
Single tracks are reconstructed in these telescopes with almost 100\% efficiency for $\pt>20\MeVc$, but because of multiple scattering and the effect of the magnetic field,
tracks can be identified as coming from the IP with an efficiency that increases as a function of \pt and is
greater than 80\% for $\pt>40\MeVc$~\cite{totemPerformance:2013}.
The minimum bias trigger, defined by the requirement of the presence of at least one track candidate in either of the T2 detectors~\cite{Antchev:2013haa}, has an efficiency close to 100\%~\cite{Chatrchyan:2014qka} for events where a charged particle is produced within the T2 acceptance.
According to the \PYTHIA8 and \textsc{qgsjet}II-04 \cite{Ostapchenko:2010vb} generators, about 91--96\%
of the total inelastic cross section at $\sqrt{s} = 8$\TeV is seen by T2 \cite{Antchev:2013paa},  with the uncertainty coming mainly from low mass diffractive events. The present analysis follows the procedure described in Ref.~\cite{Chatrchyan:2014qka}, where more details are given on the trigger, data selection, and correction procedures.

Corrections for the contribution of background events triggered by T2 but without a charged primary particle in the T2 acceptance are estimated with simulated events from \PYTHIA8 and \EPOS. These models were found
to enclose the measured pseudorapidity distributions of charged particles in the forward region~\cite{Chatrchyan:2014qka}.
The average corrections for the two models vary from 4\% and 1\% at $\ptmin \approx 1$\GeV to 7\% and 5\% at $\ptmin \approx 45$\GeV, for the track and track-jet analysis, respectively. The deviation of \PYTHIA8 and \EPOS from the average correction is taken as an estimate of the systematic uncertainty related to the T2 trigger efficiency; it is less than 0.7\% for the leading track measurement and varies between 0.1 and 1.0\% for the leading track-jet measurement~\cite{Chatrchyan:2014qka}.

Events are selected offline by requiring the presence of a leading track in the region $\abs{\eta} < 2.4$ with $\pt > 0.4$\GeV.  These events are used to normalise the integrated distributions in both the leading track and the track-jet measurements. Track-jets are reconstructed offline from tracks with $\pt > 0.1$\GeV and $\abs{\eta} < 2.4$, clustered by using the anti-\kt algorithm~\cite{Cacciari:2005hq,Cacciari:2008gp,Cacciari:2011ma} with a distance parameter of 0.5. The track-jet momentum is determined from the sum of all track momenta in the track-jet. The pseudorapidity restriction  $\abs{\eta^\text{jet}} < 1.9$ assures that the track-jet is contained within the tracker acceptance.

Detailed MC simulations of the CMS and T2 detectors are based on \GEANTfour~\cite{Agostinelli:2002hh}. Simulated events are processed and reconstructed in the same manner as collision data. For the correction of detector effects, as well as for comparison with models, both the \PYTHIA6~\cite{Sjostrand:2006za} (version 6.426) event generator with tune Z2*~\cite{Chatrchyan:2013gfi}
and the \PYTHIA8 (version 8.153) generator with tune 4C are used. The final correction is obtained by averaging those from the two generators.

The data are corrected to the stable-particle level, which is defined to include primary charged particles with lifetimes of $c\tau >1\unit{cm}$, either directly produced in the $\Pp\Pp$ collisions or from decays of particles with shorter lifetimes. According to this definition, $\PKzS$ and $\PgL$ hadrons are considered stable. Generated events are selected at the stable-particle level if at least one charged particle with $\pt >40\MeV$ is present within the range $5.3 < \abs{\eta} < 6.5$, and at least one charged particle with  $\pt >0.4\GeV$ is found within $\abs{\eta} < 2.4$.
In each event, the highest-\pt charged particle within  $\abs{\eta} < 2.4$ and  $\pt >0.8$\GeV is selected as the leading particle.
Charged particles are clustered into jets by using the anti-\kt algorithm
with a distance parameter of 0.5 with no restriction on $\pt$ or $\eta$.
The leading charged-particle jet is then defined as the charged-particle jet with the highest \pt above 1\GeV and $\abs{\eta^\text{jet}} < 1.9$.

The average systematic uncertainty in the track reconstruction efficiency is taken to be 3.9\%~\cite{CMS:2010mua}. Its effect is studied by randomly rejecting 3.9\% of the tracks and then repeating the analysis.  In the jet analysis, for tracks with low $\pt$, the rejection probability is taken as 15\% for $\pt < 1$\GeV. However, since the measurement is integrated over $\pt$, it is nearly insensitive to even such large values of the rejection probability. The resulting uncertainty varies between 0.4\% and 3.7\% for the leading charged particle analysis and between 2\% and 12\% for the leading jet analysis. The larger uncertainties correspond to higher \ptmin.

The \pt distribution of leading track-jets is unfolded to the stable-particle level by applying the iterative procedure~\cite{DAgostini:1994zf} implemented in \textsc{RooUnfold}~\cite{Adye:2011fk} in order to correct for the jet reconstruction efficiency and for migrations in jet \pt. Thanks to the good \pt resolution of the reconstructed tracks a simple correction for the track-finding efficiency is found to be sufficient for obtaining the \pt distribution of leading charged particles.
The \PYTHIA6 and \PYTHIA8 MC models are used to generate the response matrices and efficiency corrections, and the average correction from the two generators is used to obtain the \pt distributions at stable-particle level.
The corrections vary between 5\% and 10\% at $\pt \approx 1$\GeV, to 10\% and 40\% at $\pt \approx 45$\GeV, for the charged particle and the jet measurements, respectively. The deviation from the average is taken as an estimate of the systematic uncertainty related to the correction procedure. This uncertainty varies from 0.6 to 3\% for the leading charged particle analysis, and from 2 to 10\% for the leading jet analysis, depending on  \ptmin.

The systematic uncertainties are summarised in Table~\ref{tab:sys}.

\begin{table}[htb]
\centering
\topcaption{The systematic uncertainties for the leading charged particle ($0.8 < \ptmin < 50$\GeV) and leading jet ($1 < \ptmin <  50 $\GeV) measurements.}
\label{tab:sys}
\begin{tabular}{ccc}
\hline
\multirow{2}{*}{Source} & \multicolumn{2}{ c }{\qquad Uncertainty (\%)}  \\
\cline{2-3}
 & Leading charged particle & Leading jet  \\
\hline
T2 trigger efficiency                & 0.7  & 0.1--1.0 \\
Tracking efficiency          & 0.4--3.7   &   2--12   \\
Correction procedure             & 0.6--3.0  &   2.0--10  \\
\hline
Total                        & 0.7--4.6 & 2.5--16  \\
\hline
\end{tabular}
\end{table}

The per-event yields, defined in  Eq.~(\ref{int}), are obtained experimentally as
\begin{equation}
r(\ptmin)
= \frac {1}{N_{\text{evt}}} \sum_{\ptlead>\ptmin} \Delta \ptlead
\left(\frac{\Delta N}{\Delta \ptlead}\right),
\end{equation}
where $N_{\text{evt}}$ is the number of events with a leading charged particle within $\abs{\eta} < 2.4$ and with $\pt > 0.4$\GeV, $\Delta \ptlead$ is the bin width, and $\Delta N$ is the number of events with a leading charged particle or leading jet in the bin.

\section{Results}
\begin{figure*}[htbp!]
  \centering
    \includegraphics[width=0.485\textwidth]{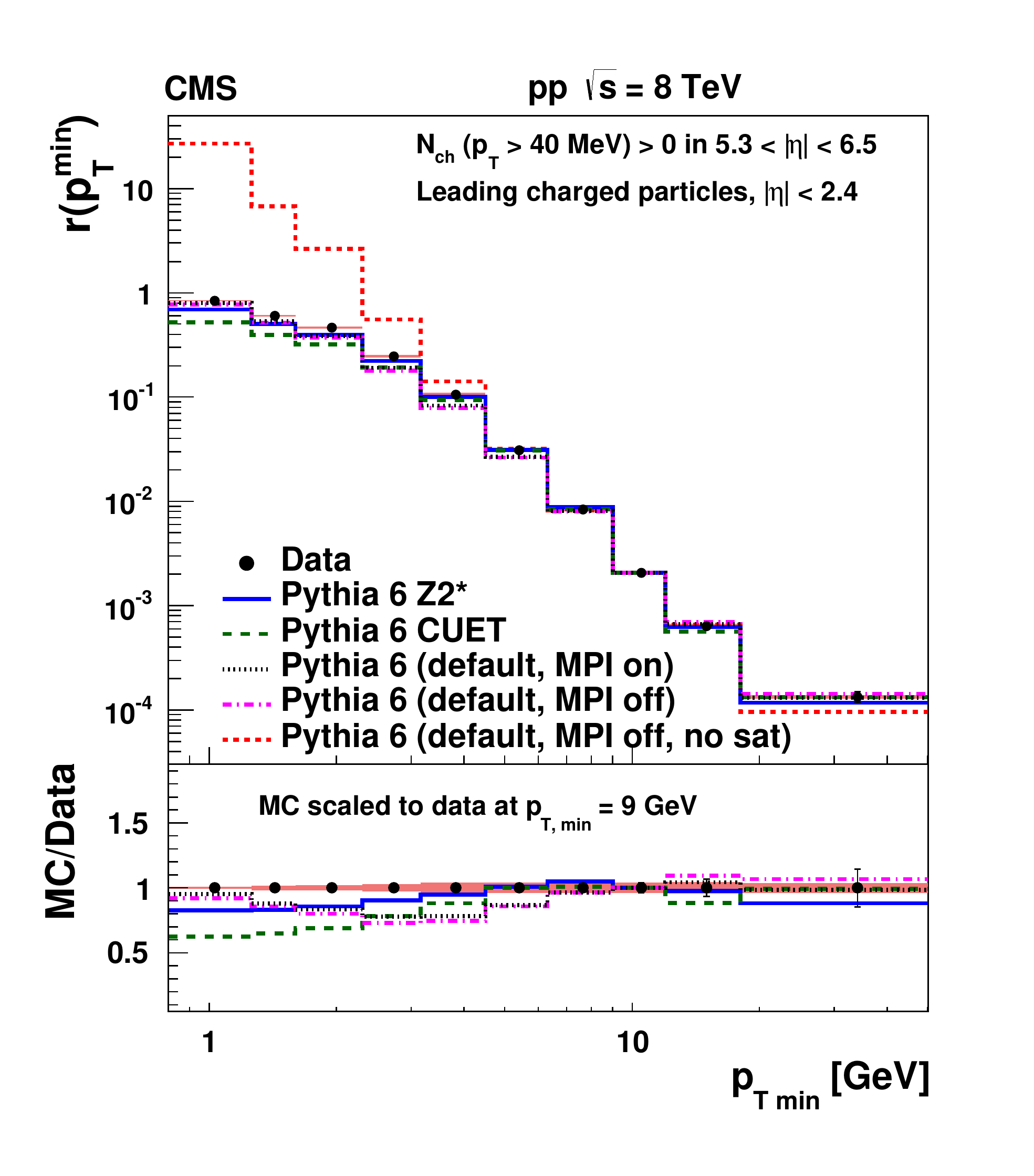}
    \includegraphics[width=0.485\textwidth]{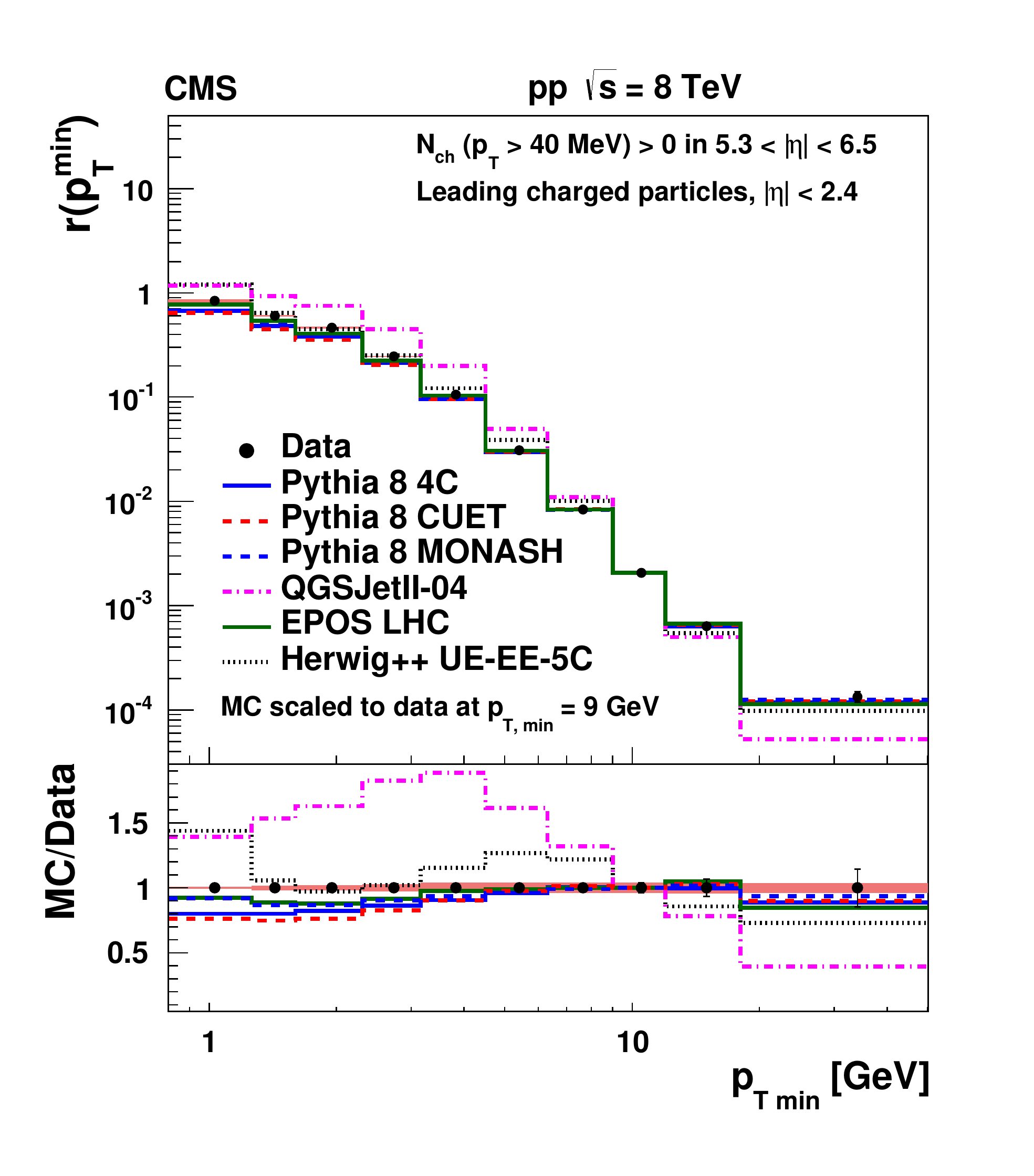}
    \includegraphics[width=0.485\textwidth]{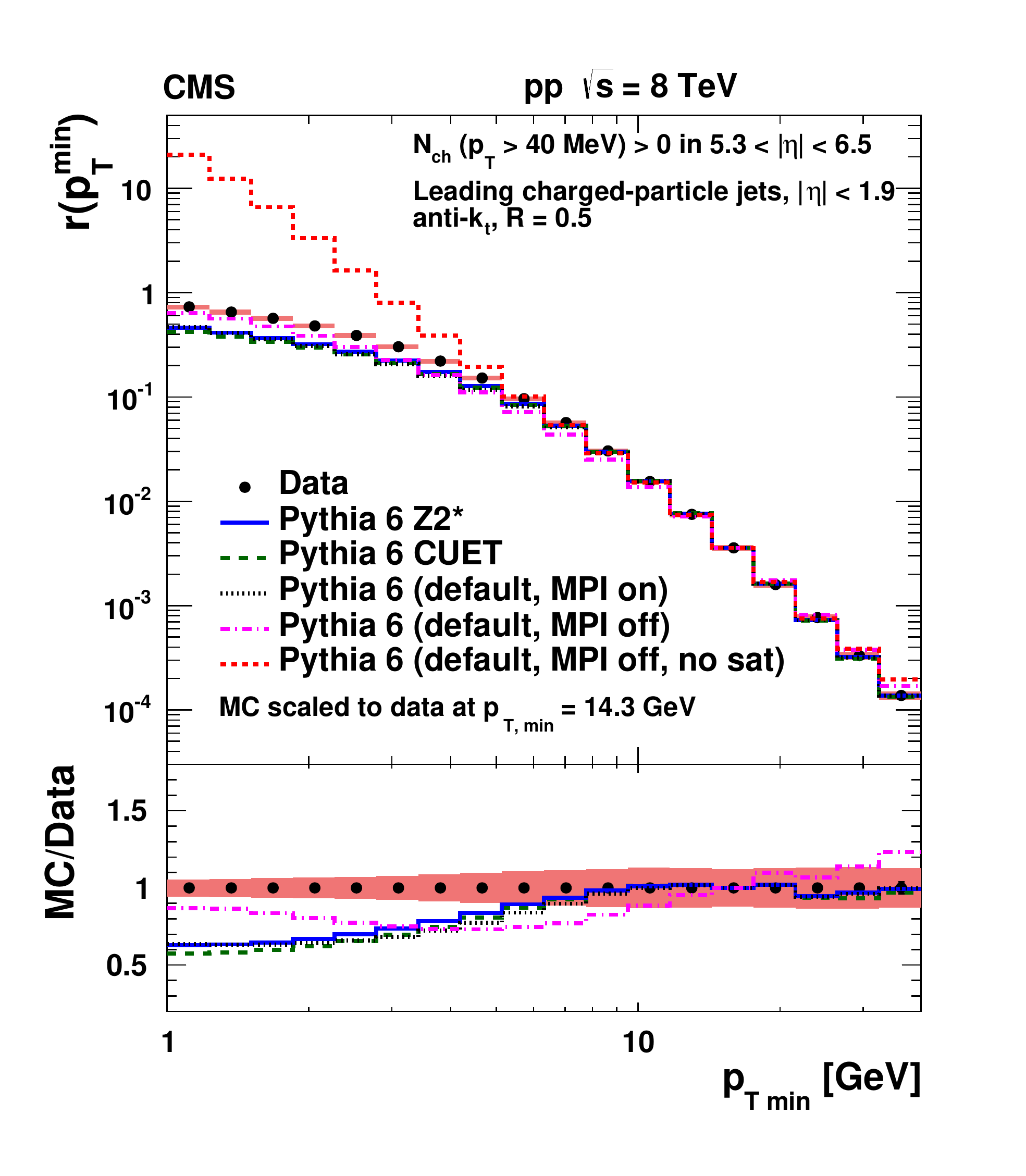}
    \includegraphics[width=0.485\textwidth]{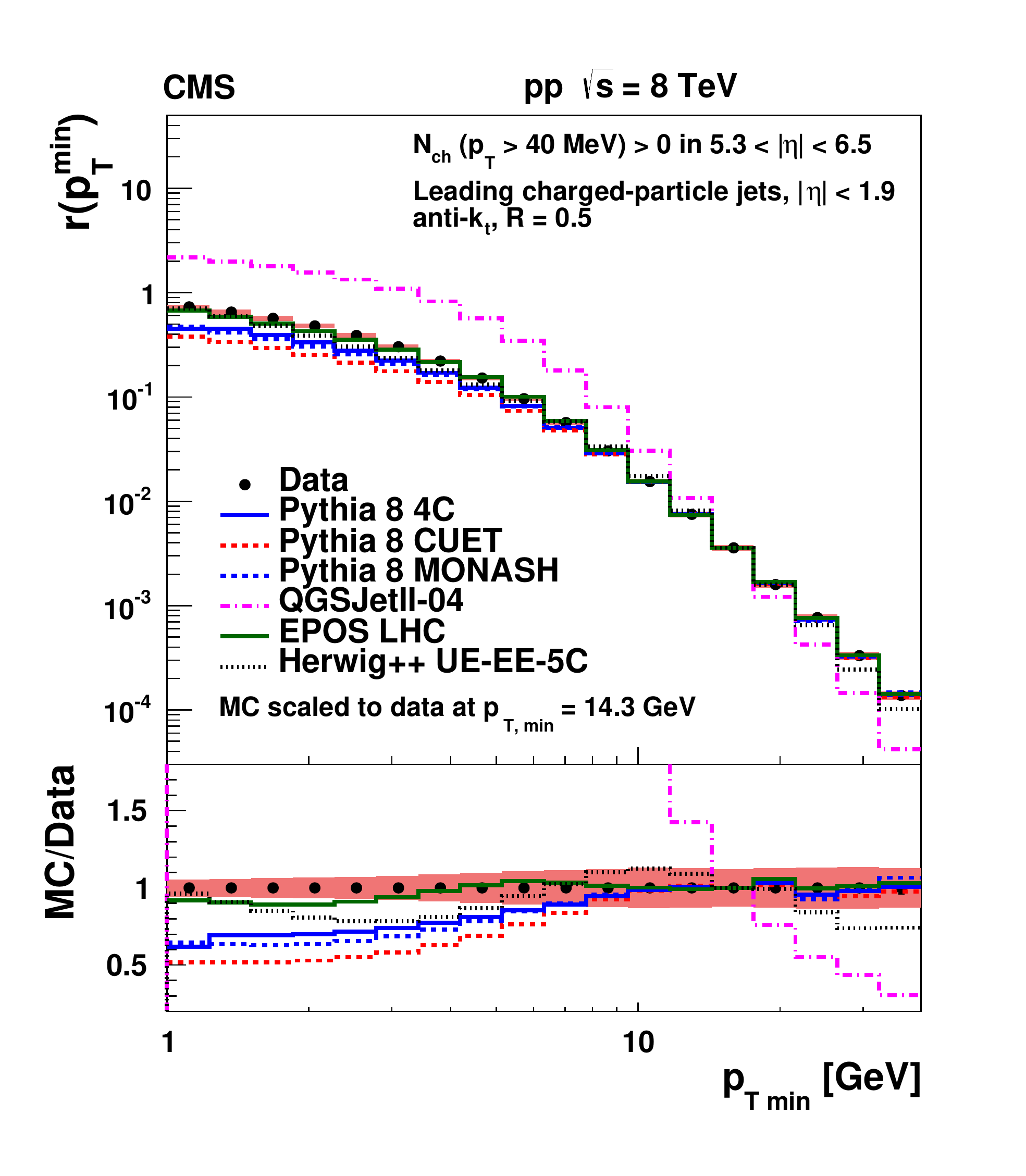}
  \caption{The integrated yield, $r(\ptmin)$, of events with a leading charged particle within $\abs{\eta} < 2.4$ (top) and with a leading jet within $\abs{\eta} < 1.9$ (bottom), as a function of \ptmin. The data are compared to predictions from several \PYTHIA6 tunes (left) and various
other event generators (right). The lower panels show the ratios of the MC and the data yields (MC/Data). The error bars indicate the statistical uncertainty and the red shaded area (only visible in the ratio plots) represents  the systematic uncertainty. The predictions are scaled to the measured value of $r(\ptlead > 9.0\GeV$) (top) and $r(\ptlead > 14.3\GeV)$ (bottom).
The prediction from \PYTHIA6 with MPI off and no parton saturation is not shown in the MC/data ratio plot (left) because of the large disagreement with the data.
}
 \label{fig:tracks}
\end{figure*}

Figure~\ref{fig:tracks}
shows the integrated distributions for the leading charged particle and leading jet events for $\ptmin > 0.8$ and 1\GeV, respectively. The distributions fall steeply at large transverse momenta, and by construction approach unity at small \ptmin.
The turnover from a relatively flat to a steeply-falling distribution takes place between 1 and
10\GeV. However, the turnover point is different for the leading charged particles and the leading jet measurements. This reflects the fact that when particles are clustered into jets, more energy from additional particles is collected within the jet cone. In fact, when the jet cone size is reduced, the leading jet distribution approaches the leading charged particle distribution.

For the comparison of the data to predictions of QCD MC generators the latter are rescaled to describe the high-\ptlead region. This rescaling is applied because the normalisation to the total visible cross section, which depends on the low-\pt regularisation, affects the values of $r$ also at high-\ptlead, where in fact theoretical predictions are more robust and agree better with the data. The exact choice of the normalisation point is arbitrary---$r(\ptlead > 9.0\GeV)$ for the leading charged particle, and $r(\ptlead > 14.3\GeV)$ for the leading jet---and the conclusions from this study are drawn from the shape of the distributions alone. The predictions at small \ptlead thus give information on the modelling of the transition region from large to small \ptlead.

In Fig.~\ref{fig:tracks} (left plots)
the yields $r(\ptmin)$ as a function of \ptmin are compared to the predictions of the event generator \PYTHIA6 with tunes Z2* and CUET, as well as with the default version of \PYTHIA6, both with and without MPI. Also shown is the impact of turning off the regularisation of the cross section, labeled ``\PYTHIA6 (default, MPI off, no sat)".
At low \ptmin, the distribution predicted by this latter model differs by more than one order of magnitude
from predictions with the regularised cross section.

In Fig.~\ref{fig:tracks} (right plots)
the leading charged particle and leading jet data are compared with \PYTHIA8 with tunes 4C, CUET, and MONASH, \HERWIGpp (version 2.7.0) with tune UE-EE-5C, \EPOS (version 1.99) with LHC tune, and \textsc{qgsjet}II-04.

The leading charged particle and leading jet cross sections are best described by \EPOS, which deviates only by up to 10\% from the data at very low $\ptmin$ and reproduces the data well for $\ptmin>4$\GeV. The event generator \HERWIGpp (UE-EE-5C tune) describes
the leading jet cross sections fairly well, but does not reproduce the transition from large to small \pt in the leading charged particle
cross section.
The event generators \PYTHIA6 (Z2* and CUET tunes) and \PYTHIA8 (4C, CUET, and MONASH tunes) predict
a somewhat different shape for the measured distributions at small \pt.

The comparison of the MC predictions for MPI switched on and off indicates that the effect of MPI is small for leading charged particles, since the particle multiplicity plays only a minor role.
However, when clustering particles into jets, the additional particles from MPI play a role, and a large difference is seen when such interactions are switched off in the simulation as in
Fig.~\ref{fig:tracks} (bottom left); this brings \PYTHIA6 closer to the data at low \ptmin.

The predictions with MPI and saturation turned off (dashed curves in Fig.~\ref{fig:tracks}, left plots) exhibit a significant deviation from the data at small \pt.

In general,  \PYTHIA and \HERWIGpp describe the trend of the measured distributions but fail to reproduce the details in the $\mathcal{O}$(1--5\GeV) region, which calls for an improvement in their modelling of the transition from the non-perturbative to perturbative regime.

\section{Summary}

The integrated yields of events with a leading charged particle or a leading charged-particle jet with \pt above a given $\ptmin$ threshold, starting at $\ptmin =0.8$ and 1\GeV, respectively, have been measured
in $\Pp\Pp$ collisions at $\sqrt{s}=8$\TeV in a data sample corresponding to an integrated luminosity of 45\mubinv. The particles and jets are measured in the pseudorapidity ranges
$\abs{\eta} < 2.4$ and 1.9, respectively.

The yields are found to be relatively flat in the \ptmin region around 1\GeV---where the fixed-order perturbative parton-parton cross section diverges in the absence of any mechanism that saturates or unitarises the pQCD scattering---followed by a steep decrease for $\ptmin>10$\GeV. The flattening behaviour observed at very low \ptmin is best described by \EPOS, which deviates by at most 10\% from the data. The comparison of the data with different phenomenological predictions of hadronic interaction models may help to improve the description of the transition between the perturbative and non-perturbative QCD regimes,
which is dominated by the effects of parton density saturation and multiple partonic interactions.

\begin{acknowledgments}
We congratulate our colleagues in the CERN accelerator departments for the excellent performance of the LHC and thank the technical and administrative staffs at CERN and at other CMS institutes for their contributions to the success of the CMS effort. We are very grateful to the TOTEM Collaboration for making their trigger signal available to CMS and for providing the additional tools required to analyse the acquired data. In addition, we gratefully acknowledge the computing centres and personnel of the Worldwide LHC Computing Grid for delivering so effectively the computing infrastructure essential to our analyses. Finally, we acknowledge the enduring support for the construction and operation of the LHC and the CMS detector provided by the following funding agencies: BMWFW and FWF (Austria); FNRS and FWO (Belgium); CNPq, CAPES, FAPERJ, and FAPESP (Brazil); MES (Bulgaria); CERN; CAS, MoST, and NSFC (China); COLCIENCIAS (Colombia); MSES and CSF (Croatia); RPF (Cyprus); MoER, ERC IUT and ERDF (Estonia); Academy of Finland, MEC, and HIP (Finland); CEA and CNRS/IN2P3 (France); BMBF, DFG, and HGF (Germany); GSRT (Greece); OTKA and NIH (Hungary); DAE and DST (India); IPM (Iran); SFI (Ireland); INFN (Italy); MSIP and NRF (Republic of Korea); LAS (Lithuania); MOE and UM (Malaysia); CINVESTAV, CONACYT, SEP, and UASLP-FAI (Mexico); MBIE (New Zealand); PAEC (Pakistan); MSHE and NSC (Poland); FCT (Portugal); JINR (Dubna); MON, RosAtom, RAS and RFBR (Russia); MESTD (Serbia); SEIDI and CPAN (Spain); Swiss Funding Agencies (Switzerland); MST (Taipei); ThEPCenter, IPST, STAR and NSTDA (Thailand); TUBITAK and TAEK (Turkey); NASU and SFFR (Ukraine); STFC (United Kingdom); DOE and NSF (USA).

Individuals have received support from the Marie-Curie programme and the European Research Council and EPLANET (European Union); the Leventis Foundation; the A. P. Sloan Foundation; the Alexander von Humboldt Foundation; the Belgian Federal Science Policy Office; the Fonds pour la Formation \`a la Recherche dans l'Industrie et dans l'Agriculture (FRIA-Belgium); the Agentschap voor Innovatie door Wetenschap en Technologie (IWT-Belgium); the Ministry of Education, Youth and Sports (MEYS) of the Czech Republic; the Council of Science and Industrial Research, India; the HOMING PLUS programme of the Foundation for Polish Science, cofinanced from European Union, Regional Development Fund; the Compagnia di San Paolo (Torino); the Consorzio per la Fisica (Trieste); MIUR project 20108T4XTM (Italy); the Thalis and Aristeia programmes cofinanced by EU-ESF and the Greek NSRF; the National Priorities Research Program by Qatar National Research Fund; and Rachadapisek Sompot Fund for Postdoctoral Fellowship, Chulalongkorn University (Thailand).
\end{acknowledgments}

\bibliography{auto_generated}

\providecommand{\href}[2]{#2}\begingroup\raggedright\begin{thebibliography}{10}%
\makeatletter
\providecommand{\hrefCMSnoop }[0]{\@secondoftwo}%
\makeatother
\providecommand{\doi}{\texttt{doi:}\begingroup \urlstyle{tt}\Url}

\bibitem{Aad:2011fc}
\hrefCMSnoop {}{{ATLAS Collaboration}, ``{Measurement of inclusive jet and
  dijet production in pp collisions at $\sqrt{s}$ = 7 TeV using the ATLAS
  detector}'',} \textit{ Phys. Rev. D} \textbf{ 86} (2012) 014022,
  \href{http://dx.doi.org/10.1103/PhysRevD.86.014022}{\doi{10.1103/PhysRevD.86.014022}},
\href{http://www.arXiv.org/abs/1112.6297}{\texttt{arXiv:1112.6297}}.

\bibitem{Chatrchyan:2012bja}
\hrefCMSnoop {}{{CMS Collaboration}, ``{Measurements of differential jet cross
  sections in proton-proton collisions at $\sqrt{s}=7$ TeV with the CMS
  detector}'',} \textit{ Phys. Rev. D} \textbf{ 87} (2013) 112002,
  \href{http://dx.doi.org/10.1103/PhysRevD.87.112002}{\doi{10.1103/PhysRevD.87.112002}},
\href{http://www.arXiv.org/abs/1212.6660}{\texttt{arXiv:1212.6660}}.

\bibitem{CMS:2011ab}
\hrefCMSnoop {}{{CMS Collaboration}, ``{Measurement of the Inclusive Jet Cross
  Section in pp Collisions at $\sqrt{s}$ = 7 TeV}'',} \textit{ Phys. Rev.
  Lett.} \textbf{ 107} (2011) 132001,
  \href{http://dx.doi.org/10.1103/PhysRevLett.107.132001}{\doi{10.1103/PhysRevLett.107.132001}},
\href{http://www.arXiv.org/abs/1106.0208}{\texttt{arXiv:1106.0208}}.

\bibitem{Antchev:2013paa}
\hrefCMSnoop {}{{TOTEM} Collaboration, ``{Luminosity-Independent Measurement of
  the Proton-Proton Total Cross Section at $\sqrt{s}=8$~TeV}'',} \textit{ Phys.
  Rev. Lett.} \textbf{ 111} (2013) 012001,
  \href{http://dx.doi.org/10.1103/PhysRevLett.111.012001}{\doi{10.1103/PhysRevLett.111.012001}}.

\bibitem{Sjostrand:1987su}
\hrefCMSnoop {}{T.~Sj{\"o}strand and M.~van Zijl, ``{A Multiple Interaction
  Model for the Event Structure in Hadron Collisions}'',} \textit{ Phys. Rev.
  D} \textbf{ 36} (1987) 2019,
\href{http://dx.doi.org/10.1103/PhysRevD.36.2019}{\doi{10.1103/PhysRevD.36.2019}}.

\bibitem{Sjostrand:2004pf}
\hrefCMSnoop {}{T.~Sj{\"o}strand and P.~Z. Skands, ``{Multiple interactions and
  the structure of beam remnants}'',} \textit{ JHEP} \textbf{ 03} (2004) 053,
  \href{http://dx.doi.org/10.1088/1126-6708/2004/03/053}{\doi{10.1088/1126-6708/2004/03/053}}.

\bibitem{Dischler:2000pk}
\hrefCMSnoop {}{J.~Dischler and T.~Sjostrand, ``A toy model of color screening
  in the proton'',} \textit{ Eur. Phys. J. direct} \textbf{ 3} (2001) 1,
  \href{http://dx.doi.org/10.1007/s1010501c0002}{\doi{10.1007/s1010501c0002}},
\href{http://www.arXiv.org/abs/hep-ph/0011282}{\texttt{arXiv:hep-ph/0011282}}.

\bibitem{Mueller:1989st}
\hrefCMSnoop {}{A.~H. Mueller, ``{Small-$x$ behavior and parton saturation: A
  QCD model}'',} \textit{ Nucl. Phys. B} \textbf{ 335} (1990) 115,
\href{http://dx.doi.org/10.1016/0550-3213(90)90173-B}{\doi{10.1016/0550-3213(90)90173-B}}.

\bibitem{Grebenyuk:2012qp}
A.~Grebenyuk\hrefCMSnoop {}{ {et~al.}, ``{Jet production and the inelastic $pp$
  cross section at the LHC}'',} \textit{ Phys. Rev. D} \textbf{ 86} (2012)
  117501,
  \href{http://dx.doi.org/10.1103/PhysRevD.86.117501}{\doi{10.1103/PhysRevD.86.117501}},
\href{http://www.arXiv.org/abs/1209.6265}{\texttt{arXiv:1209.6265}}.

\bibitem{Sjostrand:2006za}
\hrefCMSnoop {}{T.~Sj{\"o}strand, S.~Mrenna, and P.~Skands, ``{PYTHIA} 6.4
  physics and manual'',} \textit{ JHEP} \textbf{ 05} (2006) 026,
  \href{http://dx.doi.org/10.1088/1126-6708/2006/05/026}{\doi{10.1088/1126-6708/2006/05/026}},
\href{http://www.arXiv.org/abs/hep-ph/0603175}{\texttt{arXiv:hep-ph/0603175}}.

\bibitem{Sjostrand:2007gs}
\hrefCMSnoop {}{T.~Sj{\"o}strand, S.~Mrenna, and P.~Z. Skands, ``A brief
  introduction to {PYTHIA} 8.1'',} \textit{ Comput. Phys. Commun.} \textbf{
  178} (2008) 852,
  \href{http://dx.doi.org/10.1016/j.cpc.2008.01.036}{\doi{10.1016/j.cpc.2008.01.036}},
\href{http://www.arXiv.org/abs/0710.3820}{\texttt{arXiv:0710.3820}}.

\bibitem{Corke:2010yf}
\hrefCMSnoop {}{R.~Corke and T.~Sj{\"o}strand, ``Interleaved parton showers and
  tuning prospects'',} \textit{ JHEP} \textbf{ 03} (2011) 032,
  \href{http://dx.doi.org/10.1007/JHEP03(2011)032}{\doi{10.1007/JHEP03(2011)032}},
\href{http://www.arXiv.org/abs/1011.1759}{\texttt{arXiv:1011.1759}}.

\bibitem{Gunnellini:2014zba}
\href {http://pos.sissa.it/archive/conferences/203/100/DIS2014_100.pdf}{{CMS
  Collaboration}, ``{Measurement of four-jet production in proton-proton
  collisions at $\sqrt{s}$ = 7 TeV, UE tunes and double parton scattering}'',}
  in \textit{ XXII. International Workshop on Deep-Inelastic Scattering and
  Related Subjects}, p.~100.
\newblock Warsaw, Poland, 2014.
\newblock
[PoS DIS2014],.

\bibitem{Skands:2014pea}
\hrefCMSnoop {}{P.~Skands, S.~Carrazza, and J.~Rojo, ``{Tuning PYTHIA 8.1: the
  Monash 2013 Tune}'',} \textit{ Eur. Phys. J. C} \textbf{ 74} (2014) 3024,
  \href{http://dx.doi.org/10.1140/epjc/s10052-014-3024-y}{\doi{10.1140/epjc/s10052-014-3024-y}},
\href{http://www.arXiv.org/abs/1404.5630}{\texttt{arXiv:1404.5630}}.

\bibitem{Bahr:2008pv}
M.~B{\"a}hr\hrefCMSnoop {}{ {et~al.}, ``Herwig++ physics and manual'',}
  \textit{ Eur. Phys. J. C} \textbf{ 58} (2008) 639,
  \href{http://dx.doi.org/10.1140/epjc/s10052-008-0798-9}{\doi{10.1140/epjc/s10052-008-0798-9}},
\href{http://www.arXiv.org/abs/0803.0883}{\texttt{arXiv:0803.0883}}.

\bibitem{Bahr:2008dy}
\hrefCMSnoop {}{M.~Bahr, S.~Gieseke, and M.~H. Seymour, ``{Simulation of
  multiple partonic interactions in Herwig++}'',} \textit{ JHEP} \textbf{ 07}
  (2008) 076,
  \href{http://dx.doi.org/10.1088/1126-6708/2008/07/076}{\doi{10.1088/1126-6708/2008/07/076}},
\href{http://www.arXiv.org/abs/0803.3633}{\texttt{arXiv:0803.3633}}.

\bibitem{Herwig:2009soft}
\hrefCMSnoop {}{M.~B{\"a}hr, J.~M. Butterworth, S.~Gieseke, and M.~H. Seymour,
  ``{Soft interactions in Herwig++}'',} (2009).
\href{http://www.arXiv.org/abs/0905.4671}{\texttt{arXiv:0905.4671}}.

\bibitem{Ostapchenko:2010vb}
\hrefCMSnoop {}{S.~Ostapchenko, ``{Monte Carlo treatment of hadronic
  interactions in enhanced Pomeron scheme: I. QGSJET-II model}'',} \textit{
  Phys. Rev. D} \textbf{ 83} (2011) 014018,
  \href{http://dx.doi.org/10.1103/PhysRevD.83.014018}{\doi{10.1103/PhysRevD.83.014018}},
\href{http://www.arXiv.org/abs/1010.1869}{\texttt{arXiv:1010.1869}}.

\bibitem{Werner:2005jf}
\hrefCMSnoop {}{K.~Werner, F.-M. Liu, and T.~Pierog, ``{Parton ladder splitting
  and the rapidity dependence of transverse momentum spectra in deuteron-gold
  collisions at RHIC}'',} \textit{ Phys. Rev. C} \textbf{ 74} (2006) 044902,
  \href{http://dx.doi.org/10.1103/PhysRevC.74.044902}{\doi{10.1103/PhysRevC.74.044902}},
\href{http://www.arXiv.org/abs/hep-ph/0506232}{\texttt{arXiv:hep-ph/0506232}}.

\bibitem{Pierog:2013ria}
T.~Pierog\hrefCMSnoop {}{ {et~al.}, ``{EPOS LHC : test of collective
  hadronization with LHC data}'',} (2013).
\href{http://www.arXiv.org/abs/1306.0121}{\texttt{arXiv:1306.0121}}.

\bibitem{Gribov:1968fc}
\href {http://jetp.ac.ru/cgi-bin/dn/e_026_02_0414.pdf}{V.~N. Gribov, ``{A
  reggeon diagram technique}'',} \textit{ Sov. Phys. JETP} \textbf{ 26} (1968)
414.

\bibitem{dEnterria:2011kw}
D.~d'Enterria\hrefCMSnoop {}{ {et~al.}, ``{Constraints from the first LHC data
  on hadronic event generators for ultra-high energy cosmic-ray physics}'',}
  \textit{ Astropart. Phys.} \textbf{ 35} (2011) 98,
  \href{http://dx.doi.org/10.1016/j.astropartphys.2011.05.002}{\doi{10.1016/j.astropartphys.2011.05.002}},
\href{http://www.arXiv.org/abs/1101.5596}{\texttt{arXiv:1101.5596}}.

\bibitem{Gribov:1984tu}
\hrefCMSnoop {}{L.~V. Gribov, E.~M. Levin, and M.~G. Ryskin, ``Semihard
  processes in {QCD}'',} \textit{ Phys. Rep.} \textbf{ 100} (1983) 1,
\href{http://dx.doi.org/10.1016/0370-1573(83)90022-4}{\doi{10.1016/0370-1573(83)90022-4}}.

\bibitem{Chatrchyan:2014fea}
\hrefCMSnoop {}{{CMS Collaboration}, ``{Description and performance of track
  and primary-vertex reconstruction with the CMS tracker}'',} \textit{ JINST}
  \textbf{ 9} (2014) 10009,
  \href{http://dx.doi.org/10.1088/1748-0221/9/10/P10009}{\doi{10.1088/1748-0221/9/10/P10009}},
\href{http://www.arXiv.org/abs/1405.6569}{\texttt{arXiv:1405.6569}}.

\bibitem{Chatrchyan:2008zzk}
\hrefCMSnoop {}{{CMS Collaboration}, ``The {CMS} experiment at the {CERN}
  {LHC}'',} \textit{ JINST} \textbf{ 3} (2008) S08004,
\href{http://dx.doi.org/10.1088/1748-0221/3/08/S08004}{\doi{10.1088/1748-0221/3/08/S08004}}.

\bibitem{Chatrchyan:2014qka}
\hrefCMSnoop {}{{CMS and TOTEM Collaborations}, ``{Measurement of
  pseudorapidity distributions of charged particles in proton-proton collisions
  at $\sqrt{s} = 8$ TeV by the CMS and TOTEM experiments}'',} \textit{ Eur.
  Phys. J C} \textbf{ 74} (2014) 3053,
  \href{http://dx.doi.org/10.1140/epjc/s10052-014-3053-6}{\doi{10.1140/epjc/s10052-014-3053-6}},
\href{http://www.arXiv.org/abs/1405.0722}{\texttt{arXiv:1405.0722}}.

\bibitem{totem}
\hrefCMSnoop {}{{{TOTEM}} Collaboration, ``{The TOTEM Experiment at the CERN
  Large Hadron Collider}'',} \textit{ JINST} \textbf{ 3} (2008) S08007,
\href{http://dx.doi.org/10.1088/1748-0221/3/08/S08007}{\doi{10.1088/1748-0221/3/08/S08007}}.

\bibitem{totemPerformance:2013}
\hrefCMSnoop {}{{TOTEM} Collaboration, ``{Performance of the TOTEM Detectors at
  the LHC}'',} \textit{ Int. J. Mod. Phys. A} \textbf{ 28} (2013) 1330046,
  \href{http://dx.doi.org/10.1142/S0217751X13300469}{\doi{10.1142/S0217751X13300469}},
  \href{http://www.arXiv.org/abs/1310.2908}{\texttt{arXiv:1310.2908}}.

\bibitem{Antchev:2013haa}
\hrefCMSnoop {}{{TOTEM} Collaboration, ``{Measurement of proton-proton
  inelastic scattering cross-section at $\sqrt{s} = 7$ TeV}'',} \textit{ EPL}
  \textbf{ 101} (2013) 21003,
\href{http://dx.doi.org/10.1209/0295-5075/101/21003}{\doi{10.1209/0295-5075/101/21003}}.

\bibitem{Cacciari:2005hq}
\hrefCMSnoop {}{M.~Cacciari and G.~P. Salam, ``{Dispelling the $N^{3}$ myth for
  the $k_t$ jet-finder}'',} \textit{ Phys. Lett. B} \textbf{ 641} (2006) 57,
  \href{http://dx.doi.org/10.1016/j.physletb.2006.08.037}{\doi{10.1016/j.physletb.2006.08.037}},
\href{http://www.arXiv.org/abs/hep-ph/0512210}{\texttt{arXiv:hep-ph/0512210}}.

\bibitem{Cacciari:2008gp}
\hrefCMSnoop {}{M.~Cacciari, G.~P. Salam, and G.~Soyez, ``The
  anti-$k_{\textit{t}}$ jet clustering algorithm'',} \textit{ JHEP} \textbf{
  04} (2008) 063,
  \href{http://dx.doi.org/10.1088/1126-6708/2008/04/063}{\doi{10.1088/1126-6708/2008/04/063}},
  \href{http://www.arXiv.org/abs/0802.1189}{\texttt{arXiv:0802.1189}}.

\bibitem{Cacciari:2011ma}
\hrefCMSnoop {}{M.~Cacciari, G.~P. Salam, and G.~Soyez, ``{FastJet} user
  manual'',} \textit{ Eur. Phys. J. C} \textbf{ 72} (2012) 1896,
  \href{http://dx.doi.org/10.1140/epjc/s10052-012-1896-2}{\doi{10.1140/epjc/s10052-012-1896-2}},
\href{http://www.arXiv.org/abs/1111.6097}{\texttt{arXiv:1111.6097}}.

\bibitem{Agostinelli:2002hh}
\hrefCMSnoop {}{{GEANT4} Collaboration, ``{GEANT4}---a simulation toolkit'',}
  \textit{ Nucl. Instr. Meth. A} \textbf{ 506} (2003) 250,
\href{http://dx.doi.org/10.1016/S0168-9002(03)01368-8}{\doi{10.1016/S0168-9002(03)01368-8}}.

\bibitem{Chatrchyan:2013gfi}
\hrefCMSnoop {}{{CMS Collaboration}, ``{Study of the underlying event at
  forward rapidity in pp collisions at $\sqrt{s}$ = 0.9, 2.76, and 7 TeV}'',}
  \textit{ JHEP} \textbf{ 04} (2013) 072,
  \href{http://dx.doi.org/10.1007/JHEP04(2013)072}{\doi{10.1007/JHEP04(2013)072}},
\href{http://www.arXiv.org/abs/1302.2394}{\texttt{arXiv:1302.2394}}.

\bibitem{CMS:2010mua}
\href {http://cdsweb.cern.ch/record/1279139}{{CMS Collaboration}, ``Measurement
  of Tracking Efficiency'',} CMS Physics Analysis Summary CMS-PAS-TRK-10-002,
  2010.

\bibitem{DAgostini:1994zf}
\hrefCMSnoop {}{G.~D'Agostini, ``{A Multidimensional unfolding method based on
  Bayes' theorem}'',} \textit{ Nucl. Instr. Meth. A} \textbf{ 362} (1995) 487,
\href{http://dx.doi.org/10.1016/0168-9002(95)00274-X}{\doi{10.1016/0168-9002(95)00274-X}}.

\bibitem{Adye:2011fk}
\hrefCMSnoop {}{T.~Adye, ``Unfolding algorithms and tests using {RooUnfold}'',}
  in \textit{ {PHYSTAT} 2011 Workshop on Statistical Issues Related to
  Discovery Claims in Search Experiments and Unfolding}, H.~Prosper and
  L.~Lyons, eds., p.~313.
\newblock Geneva, Switzerland, 2011.
\newblock \href{http://www.arXiv.org/abs/1105.1160}{\texttt{arXiv:1105.1160}}.
\newblock
\href{http://dx.doi.org/10.5170/CERN-2011-006.313}{\doi{10.5170/CERN-2011-006.313}}.

\end{thebibliography}\endgroup

\cleardoublepage \appendix\section{The CMS Collaboration \label{app:collab}}\begin{sloppypar}\hyphenpenalty=5000\widowpenalty=500\clubpenalty=5000\textbf{Yerevan Physics Institute,  Yerevan,  Armenia}\\*[0pt]
V.~Khachatryan, A.M.~Sirunyan, A.~Tumasyan
\vskip\cmsinstskip
\textbf{Institut f\"{u}r Hochenergiephysik der OeAW,  Wien,  Austria}\\*[0pt]
W.~Adam, T.~Bergauer, M.~Dragicevic, J.~Er\"{o}, M.~Friedl, R.~Fr\"{u}hwirth\cmsAuthorMark{1}, V.M.~Ghete, C.~Hartl, N.~H\"{o}rmann, J.~Hrubec, M.~Jeitler\cmsAuthorMark{1}, W.~Kiesenhofer, V.~Kn\"{u}nz, M.~Krammer\cmsAuthorMark{1}, I.~Kr\"{a}tschmer, D.~Liko, I.~Mikulec, D.~Rabady\cmsAuthorMark{2}, B.~Rahbaran, H.~Rohringer, R.~Sch\"{o}fbeck, J.~Strauss, W.~Treberer-Treberspurg, W.~Waltenberger, C.-E.~Wulz\cmsAuthorMark{1}
% ...
\vskip\cmsinstskip
% ...
1:~~Also at Vienna University of Technology, Vienna, Austria\\
2:~~Also at CERN, European Organization for Nuclear Research, Geneva, Switzerland\\
% ...

\end{sloppypar}
\end{document}